\documentclass[aps, prb,reprint,superscriptaddress,twocolumn,amsmath,amssymb,floatfix]{revtex4-1}

\usepackage{color}
\usepackage{graphicx}
\usepackage{dcolumn}
\usepackage{bm}
\usepackage{bbm}
\usepackage{textcomp}
\usepackage[nice]{units}

\begin{document}
\title{Experimental analysis of single particle deformations and rotations in colloidal and granular systems}
\author{M. Roth}
\altaffiliation{Graduate School Materials Science in Mainz, Staudinger Weg 9, D-55128 Mainz, Germany}
\email[]{marcel.roth@mpip-mainz.mpg.de}
\author{M. Franzmann}
\author{M. D\textasciiacute{}Acunzi}
\author{M. Kreiter}
\author{G. K. Auernhammer}
\homepage[]{http://www.mpip-mainz.mpg.de/$\sim{}$auhammer}
\affiliation{Physics of Polymers, Max Planck Institute for Polymer Research, Ackermannweg 10, 55128 Mainz, Germany}
 
\date{\today}
\begin{abstract}
Confocal microscopy of fluorescent labeled particles has been used to study the dynamical and structural properties of colloidal and granular matter in real space. Localization algorithms allow for a fully automatized determination of the three dimensional positions and translational motions of all constituent (spherical) particles in the observed volume. Though, particle deformations or rotational motions were hardly addressed. Here we present preparation and image processing techniques to also extract the deformation and rotational state of the particles. The deformation analysis is worked out for particles with a hollow sphere-like fluorescence intensity distribution that are more sensitive to small deformations. In the second case of rotations we utilize the angle dependence of the light absorption of the incorporated dye molecules to prepare optically anisotropic particles in analogy to polarized fluorescence after photobleaching. Rotations of these particles are expressed as intensity fluctuations in the confocal images. In contrast to existing methods our techniques do not reduce the quality of the actual particle localization. They can help understanding complex reorganization processes in arrested states of colloidal and granular materials during aging or under external stimuli such as shear or compression.
\end{abstract}
\maketitle
\section{Introduction}
\label{sec:Introduction}
The mechanical state of colloidal and granular systems ranges from purely viscous suspensions to networks and concentrated systems exhibiting significant elasticity. Many parameter have an impact, including the choice of materials and sample compositions or external parameters like pressure and temperature. The dominant phenomena, such as yielding\cite{JRheo.52.649, JPhysCondMat.16.S4861} or shear thinning\cite{AdvCollInterfSci.3.111} are easily identified on a macroscopic level using diverse rheometric methods \cite{ViscPropPoly.Ferry, JRheo.50.883, JRheo.49.851, JRheo.47.303, JChemPhys.132.124702}. However, the microscopic understanding of these processes is often hindered due to a lack of appropriate experimental measurement techniques on a microscopic level. Diverse theoretical approaches and simulation methods\cite{PhysRevLett.98.038301, Luding_04, Nonlinearity.22.R101, EuropPhysJE.30.275} try to fill the gap but may show discrepancies due to the complexity of the system. \\
Involved particle diameters range from $\unit[10]{nm}$ to $\unit[10]{\mu m}$ for colloidal and from $\unit[1]{\mu m}$ to $\unit[1]{mm}$ for granular matter. These lengthscales favor optical investigation methods either in scattering geometry such as dynamic and static light scattering\cite{JChemPhys.130.134907} or in direct microscopic imaging. Some of these imaging techniques are specialized to the extraction of three-dimensional (3D) particle coordinates such as x-ray tomography\cite{NatMat.7.189,Science.326.408} which can be used for particle sizes down to few tens of micrometers. In this paper we deal with particle diameters down to $\unit[1]{\mu m}$ that can be investigated with confocal microscopy \cite{JPhysCondMatt.14.7581, JCollInterfSci.179.298, Science.287.5453}. For loose gel structures as well as densely packed glassy systems 3D particle coordinates were determined with accuracies of  about $\unit[0.05]{\mu m}$ \cite{AdvCollIntSci.136.65}. Phase transitions and coexistence regions were easily identified \cite{MolCrystLiquidCryst.409.59, PhysRevE.64.021407, PhysRevLett.96.028306}. Crystallization kinetics as well as its frustration due to irregularly shaped particles or polydispersity were also investigated \cite{EurPhysJAplPhys.44.21}. Using confocal microscopes with high speed scanning\cite{EncBioMatMedEng.Weeks} or multiple beam illumination\cite{OptExpress.14.8702} also fast dynamic processes, like \textit{e.\,g.} colloidal flows under external shear \cite{PhysRevLett.99.028301, PhysFluids.20.043306}, attracted attention. \\
In contrast, colloidal gels features only slow particle rearrangements during gravitational collapse\cite{JPhysCondMatt.12.9599} or under other external forces\cite{Langmuir.17.2918}. In this and other kinetically arrested systems motions are suppressed because acting forces compensate each other \cite{PhysRevLett.81.1841}. If such systems are further compressed the loading force is heterogeneously transmitted through the particle contacts along so called force chains \cite{JPhysCondMatt.14.2391}. The characteristics of this heterogeneous distribution of forces is currently under debate. It has been simulated\cite{EuropPhysJE.30.275} and studied experimentally, \textit{e.\,g.} via materials that show stress-induced birefringence\cite{science.269.513} in case of granular systems. However, smaller particles in the micrometer range and 3D samples are inaccessible with this method. Another approach to identify contact forces between touching particles uses soft spheres that deform under external forces \cite{PhysicaA.327.201}. In other words deformable particles can be used to sense local forces. From a practical point of view, in particular hollow spheres are suitable, since their elastic constants can be tuned by the shell thickness and choice of materials \cite{Langmuir.25.2711, MatLett.61.2560}. When stressed, these hollow spheres deform to an ellipsoid-like object. The induced anisotropy and orientation of this ellipse can be used to quantify the strength and direction of the acting force. Here we present a method to automatically determine this deformation state of all hollow particles in the system together with their 3D coordinates from confocal images. The analysis is based on an already existing algorithm to locate fully labeled spheres by Crocker, Grier\cite{JCollInterfSci.179.298} and Weeks\cite{Science.287.5453}.\\
In case of non-balanced forces complex reorganization processes come into play that change the particle assembly. It is questionable whether these rearrangements involve rolling or sliding motions of the particles \cite{JApplPhys.104.054915}. Clearly frictional forces between particles in contact are essential in this problem and already were investigated in AFM studies \cite{ChemIngSci.62.2843}. Conventional confocal microscopy cannot distinguish between rolling and sliding motions because both processes manifest in simple displacements of the particles in the confocal images. Only additional information about their rotational motion will resolve this ambiguity and help to answer basic questions like: Can rotations be enforced for rough or irregularly shaped particles? How do particle-particle interactions influence rolling and sliding?\\
Rotational motions have been studied experimentally via confocal microscopy for anisotropically shaped particles or groups of bound particles \cite{Langmuir.22.7128}. However, such grouped particles might be hardly recognizable in dense suspensions. In addition, this approach immediately limits the choice of particle shapes. In particular spherical particles cannot be used. A possibility to resolve this shortcoming are particles with anisotropic optical properties. Spatial variations of fluorescent dye concentrations or incomplete metal coatings on homogeneously labeled particles\cite{Langmuir.22.9812} are directly reflected in the microscopic images and thus can be used to unambiguously determine the rotational state of the particle. However, at the same time such variations lead to inevitable complications in the analysis of the particle coordinates and eventually an alteration of the interaction behavior. To circumvent this problem, we induce an anisotropy by selectively bleaching dye molecules with specific orientations. This method was used in (polarized) fluorescence after photobleaching [(p)FRAP]\cite{Biophys.46.787} for the investigation of the rotational diffusion of particles in solution\cite{JChemPhys.120.4517} and the incorporation of dye molecules in membranes \cite{Biophys.36.73, JPhotochemBioA.181.44}. It relies on the fact, that dye molecules preferentially absorb and emit light of specific polarization \cite{PCCP.1.4571}.\\
We introduce two new methods to get mechanical information of colloidal systems on a single particle level via confocal microscopy. Hollow spheres are used to sense local forces. A distinction between translational and rotational motions is possible with the help of optically anisotropic particles. The paper is organized as follows: After an introduction into the processing of confocal data and the localization of particles, the special case of hollow spheres is discussed. The algorithm for extracting the state of deformation is presented and tested for a sample data set. In the last part we describe the method to prepare particles with an anisotropic distribution of dye molecule orientations. Rotations of these particles manifest in variations in the total fluorescence intensity if tested with linearly polarized laser light.  

\section{Materials and experimental setup}
\label{sec:Experimental}
For the deformation analysis non-labeled polystyrene (PS) template particles ($\unit[1.97]{\mu m}$ in diameter and a polydispersity of 1.05) were coated with a $\unit[60]{nm}$ silica shell \cite{Langmuir.25.2711}. Rhodamine B was covalently incorporated into the silica matrix \cite{Langmuir.25.2711, Langmuir.10.1427}. The particles were dispersed without any further surface modification in a melt of modified PDMS (Laser liquid\texttrademark with defined refractive index of $n=1.5780\pm 0.0002$, Cargille Laboratories, USA) to match the refractive index of PS.

In case of the rotational analysis silica particles with a diameter of $\unit[780]{nm}$ (polydispersity of 1.1) were synthesized with covalently bound Rhodamine B \cite{Langmuir.10.1427}. Due to the stepwise growth of the particles the dye was incorporated only in the core of the particles while the shell is unlabeled. The particles were dried in a observation cell made of glass and infiltrated with refractive index matching mixture of glycerol and water. The system was sealed with UV-curing glue to prevent evaporation. 

\begin{figure}
\includegraphics[width=0.42\textwidth]{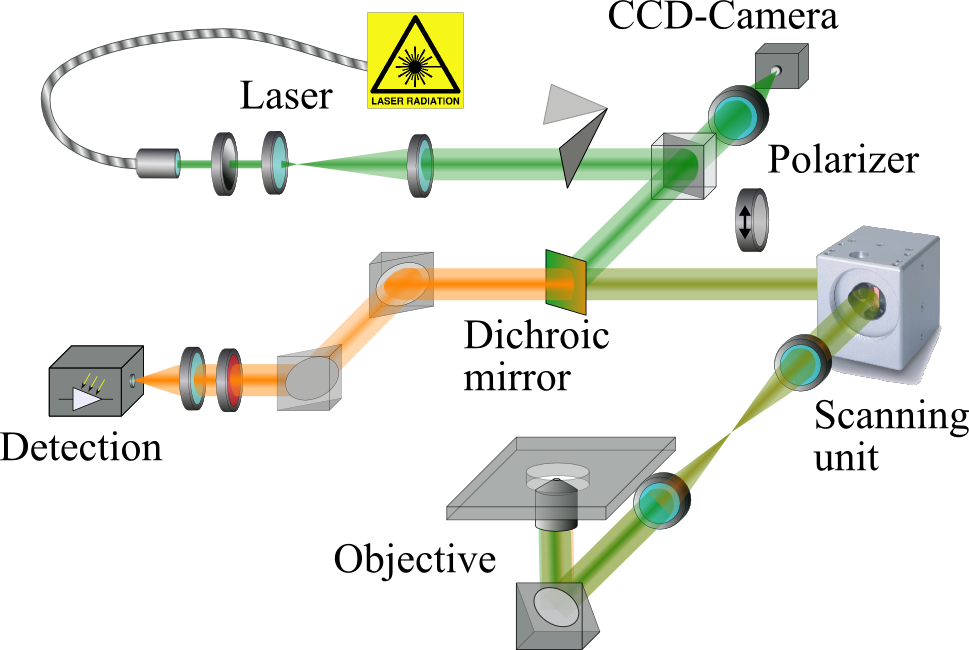}
\caption{Schematic setup of the home-made laser scanning confocal microscope. The polarizer between the dichroic mirror and the scanning unit is used only in the analysis of rotational motion.}
\label{fig:SchematicSetup}
\end{figure}

3D images were obtained with a home-made laser scanning confocal microscope (LSCM) system working in fluorescence mode [see Fig.\,\ref{fig:SchematicSetup}] \cite{Scanning.10.128}. The probing laser (Cobolt Samba\texttrademark{ }$\unit[532]{nm}$, $\unit[25]{mW}$, Cobolt AB, Sweden) was focused with a $\unit[100]{\times}$ oil immersion objective (Olympus, UPlanApo PH3, NA=1.35) into the sample and locally excited the contained dye molecules. A galvanometer based scanning unit (SCANcube$^\text{\textregistered}$ 7, SCANLAB AG, Germany) was used to scan the focus across the fixed sample in x,y-directions at a maximum speed of about two 2D-frames per second. The z-position of the focus was controlled by moving the objective with a piezo-positioning system (nanoX 200 S, ENV 40 SG nanoX, Piezosysteme Jena GmbH, Germany). The emitted fluorescent light [Rhodamin B (9-(2-Carboxyphenyl)-3,6-bis(diethylamino)xanthyliumchlorid)
] was collected by the same focusing objective and separated from the excitation laser light firstly by a dichroic mirror (Laser-beamsplitter z 532 RDC, 90\% transmission above $\unit[558]{nm}$, AHF Analysetechnik AG, Germany) and directly before the detector by a laser clean-up filter (AFH Analysetechnik AG, Germany). An avalanche photo diode (id100, idquantique, Switzerland) was used for detection at a maximum count rate of $\unit[17]{MHz}$. Since single photons were registered with high quantum efficiency only small input intensities of down to $\unit[50]{nW}$ were needed.  For the polarization dependent bleaching of the sample a thin film polarizer (VIS 4 K, suppression ratio 1:4000, Linos Photonics GmbH \& Co. KG, Germany) was placed between the dichroic mirror and the scanning unit.

\section{Basics of particle localization}
\label{sec:BasicLocal}
As the laser is scanned across the sample the detected fluorescence intensity maps the distribution of dye molecules inside the sample. If the dye molecules are incorporated into particles with a local distribution $g_\text{dye}(\textbf{r})$ the detected fluorescence intensity $I_d$ at point $\textbf{r}=(x,y,z)$ is proportional to
\begin{eqnarray}
I_d(\textbf{r}) \sim \int\limits_{V} g_\text{PSF}(\textbf{r}-\textbf{r}^\prime)\,\left(\sum\limits_{n}  g_\text{dye}(\textbf{r}^\prime-\textbf{r}_n)\right)\,d^3r^{\prime}
\label{eqn:fluodetect}
\end{eqnarray}
$\sum_n$ denotes the sum over all particles $n$ at their respective positions $\textbf{r}_n$ in the observation volume. The convolution of $g_\text{dye}$ with the so called point spread function\cite{HandbookLaserOptics_WaveOptics} $g_\text{PSF}$ accounts for the finite spatial resolution of the imaging process due to the diffraction limitation of light \cite{HandbookCM_FundamentalLimits}. The aim of any localization procedure is to extract the particle positions $\textbf{r}_n$ from the measured fluorescence intensity. Iterative deconvolution methods\cite{Meth.19.373} can be used to solve equation \eqref{eqn:fluodetect} for $\textbf{r}_n$ if the distributions $g_\text{dye}$ and $g_\text{PSF}$ are known. While $g_\text{PSF}$ can be obtained theoretically\cite{HandbookLaserOptics_WaveOptics} and in experiments\cite{ApplPhysLett.90.031106}, $g_\text{dye}$ is accessible via reasonable assumptions. However, these algorithms suffer from long computation times. 

\begin{figure}
\includegraphics[width=0.43\textwidth]{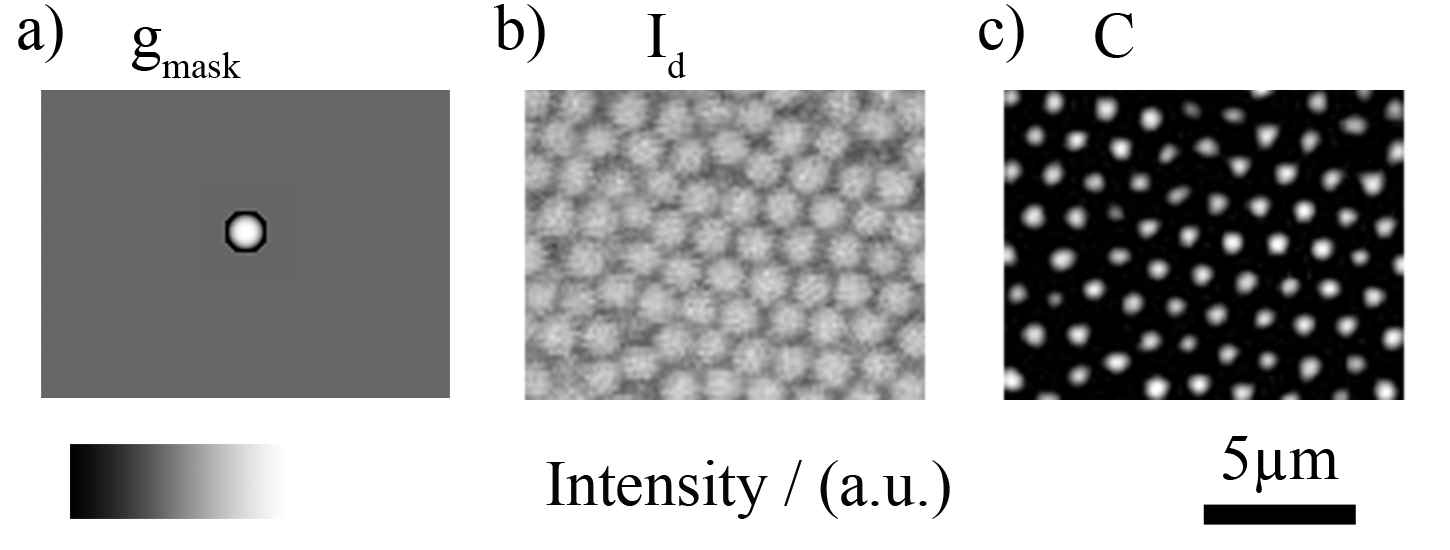}
\caption{Filtering procedure: b) In the initial confocal image the particles are represented by broad maxima at a low signal-to-noise ratio. After convolution of the intensity data with a background corrected cut Gaussian mask distribution, shown in a), the maxima are well separated and clearly identifiable in c).  }
\label{fig:LocalizePart}
\end{figure}

The problem is simplified for big and/or well separated particles for which $g_\text{PSF}$ only marginally affects the appearance of two neighboring particles compared to isolated ones. The convolution integral and the summation in equation \eqref{eqn:fluodetect} can be commuted with only small errors and the apparent distribution function 
\begin{eqnarray}
g_\text{app}(\textbf{r}-\textbf{r}_n)=\int\limits_{V}  
g_\text{PSF}(\textbf{r}-\textbf{r}^\prime)\, g_\text{dye}(\textbf{r}^\prime-\textbf{r}_n) \,d^3r^{\prime}
\end{eqnarray} 
defines the appearance of a single particle in the confocal image. That is, the particles are represented by more or less pronounced and separated intensity maxima like in Fig.\,\ref{fig:LocalizePart}\,b). In this case a time-expensive deconvolution procedure is unnecessary and generally is replaced by a more practical approach \cite{OptExpress.14.8702}. In an algorithm introduced by Crocker and Grier\cite{JCollInterfSci.179.298} and extended by Weeks\cite{Science.287.5453} a local filtering is used to improve the image quality and thus the actual particle localization. This filtering is done via an additional 3D-convolution of the experimental data set with a mask distribution $g_\text{mask}$:
\begin{eqnarray}
C(l,m,n) = \sum\limits_{i,j,k}g_\text{mask}({\scriptstyle l-i},{\scriptstyle m-j},{\scriptstyle n-k}) \, I_d(i,j,k)
\end{eqnarray}
In this step we switched from continuous coordinates $\textbf{r}$ to discrete indexes $i,\,j,\,k$ to account for the nominal spatial resolution of the experimental measurement. This set of indexes denote the position of the volume element at which the fluorescence intensity $I_d(i,j,k)$ is recorded. In our case the indexes refer to an orthogonal, equally spaced 3D grid. 
If the discrete mask distribution $g_\text{mask}$ [Fig.\,\ref{fig:LocalizePart}\,a)] is modeled to match the apparent distribution $g_\text{app}$ the 3D convolution intensity $C$ is maximal in the center of the particle. Here the measured local fluorescence intensity and the mask distribution overlap completely. With increasing separation of mask and real particle image the convolution intensity decreases. This procedure not only sharpens the intensity maxima representing the particles but also reduces data scatter by replacing $I_d$ with the locally averaged quantity $C$ [Fig.\,\ref{fig:LocalizePart}\,c)]. Now the particles can be easily localized by finding local maxima in $C$. An additional local background subtraction further simplifies the localization process. This can be incorporated into $g_\text{mask}$ by subtracting the reciprocal of its total sum:
\begin{eqnarray}
\nonumber\tilde{g}_\text{mask}(i,j,k) &=& g_\text{mask}(i,j,k) \\
 &&\qquad - 1/\left[\sum\limits_{lmn} g_\text{mask}(l,m,n)\right]
\label{eqn:masknorm}
\end{eqnarray}

\section{Results}
\label{sec:Results}
\subsection{Localization of hollow particles}
In the following we will describe the peculiarities of localizing a particle with a hollow sphere intensity distribution \footnote{In the present study we used solid particles with a labeled shell to guarantee a spherical shape of the particles. Hollow particles are prone to plastic deformation and therefore should not be used as a reference sample.}. We found that by modeling $g_\text{mask}$ to the apparent distribution $g_\text{app}$ the above described filtering process can be used for particles with arbitrary spherical symmetric profiles, i.e. also for hollow spheres. Also ellipsoid particles with moderate aspect ratios are accessible. The latter case will be important for the deformation analysis in the following section. On the contrary anisotropically shaped particles without intrinsic mirror symmetries cannot be localized reliably. 

\begin{figure}
\includegraphics[width=0.42\textwidth]{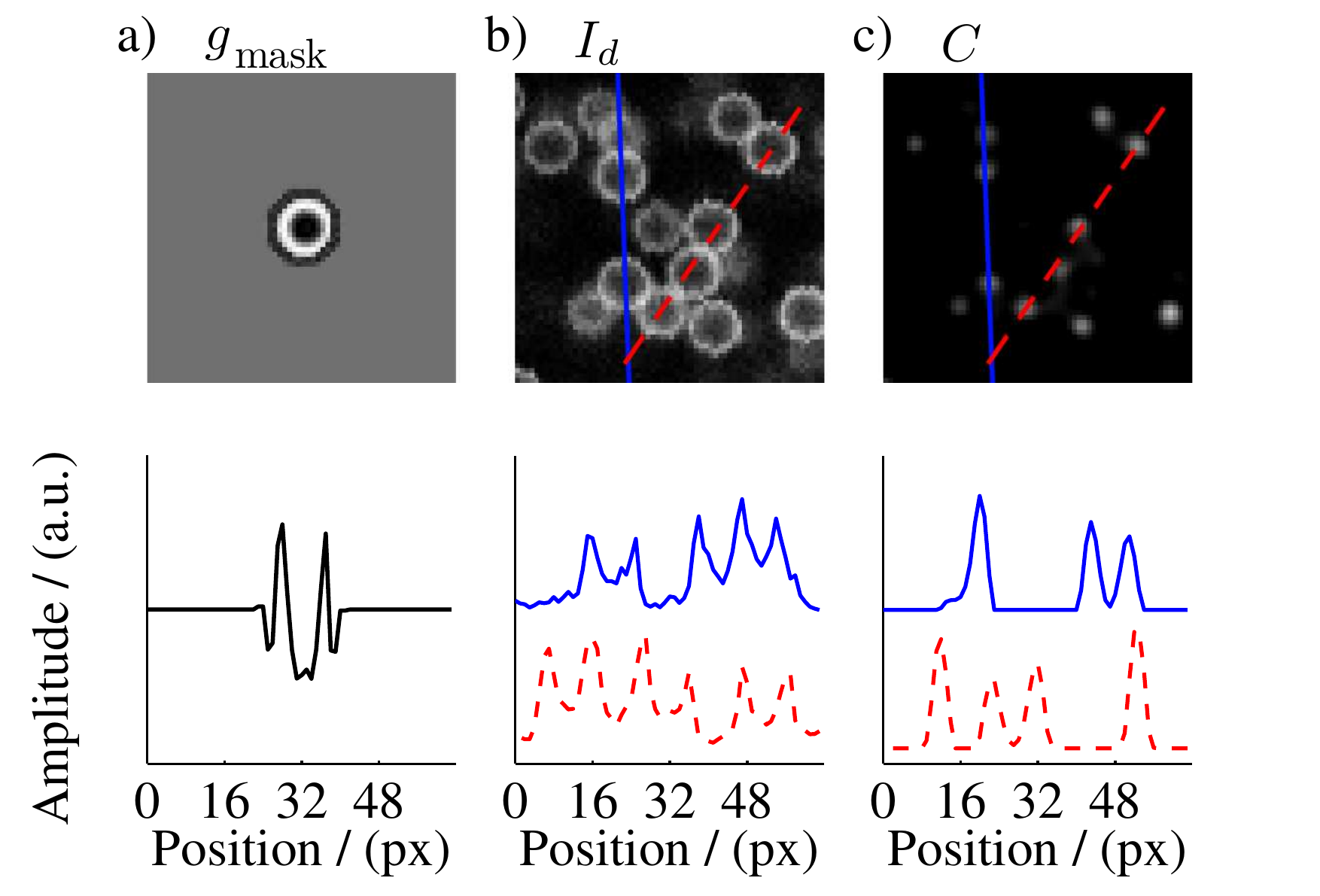}
\caption{xy-slice of the hollow sphere mask a), raw $I_d$ b) and convolved $C$ c) 3D fluorescence intensity of hollow spheres. The color scale is linear and in arbitrary units. The line profiles in the lower part of the images cross several particles. Even small deviations of the line from the center of the particle lead to significant reduction of the convolution intensity for both, lateral (x,y) and vertical (z) directions. Hence, separation distances are large in the filtered data set and positioning is easy and reliable.}
\label{fig:SampleDataCorrelMap}
\end{figure}

Although a fast implementation of the algorithm is already available in IDL \footnote{http://www.physics.emory.edu/$\sim$weeks/idl, E. Weeks} and Matlab computing languages, the code for 3D-convolution was rewritten in IDL using 3D-fast Fourier transform calculation methods \footnote{We will publish the complete code to determine the deformation of the hollow particle on our homepage.}. The original implementation can only cope with Gaussian intensity profiles since the 3D convolution is separated into three one-dimensional calculations. So, in particular measured intensity profiles that have a maximum at the rim of the particles cannot be handled. The full 3D calculation in our implementation of the convolution integral does not impose any restrictions to the mask profile $g_\text{mask}$ but requires increased memory capacities and therefore is limited to smaller data sets. However, the total computational afford does not change significantly if the whole 3D data set is split up into sufficiently small fractions in order to reduce memory occupation. The coordinate data is recombined in the end. The part of the implementation for the final localization of the particles was not changed. The algorithm was tested for real data sets obtained with our confocal microscope described above. Hence, spatial resolution, background level and optical artifacts of the confocal images like e.g. in Fig.\,\ref{fig:SampleDataCorrelMap}\,b) were representative.

As an example of the general performance of the filtering procedure we present in Fig.\,\ref{fig:SampleDataCorrelMap} b) a 64$\times$64 pixel$^2$ sized subsection of a xy-slice of a complete 3D data set. The intensity mask $g_\text{mask}$ in Fig.\,\ref{fig:SampleDataCorrelMap}\,a) was modeled according to the appearance of the particles in the image. The convolved image in Fig.\,\ref{fig:SampleDataCorrelMap}\,c) revealed well separated maxima at the hollow sphere positions that could be easily localized. Two aspects are worth mentioning: Due to the mask normalization in equation \eqref{eqn:masknorm} and a lower cutoff at zero for the convolution intensity $C$ the background was strongly reduced. Second, the maxima in the convolved image came out sharp with extensions comparable to the thickness of the particle shells in the original image. This becomes more obvious when looking at the intersections in Fig.\,\ref{fig:SampleDataCorrelMap}\,b) that do not cross all particles perfectly in the middle plane. Already small lateral (x,y) and/or vertical (z) deviations from the center lead to a significant reduction of convolution intensity. Moreover, the profiles in Fig.\,\ref{fig:SampleDataCorrelMap}c) show less noise compared to the original data aiding the subsequent localization process.\\
Any variation of the measured intensity distribution to the modeled mask profile $g_\text{mask}$ gave rise to a broadened and distorted convolution signal. If the deviations, \textit{e.\,g.} in particle size, were too large the maximal convolution intensity was no longer located in the center of the particle but still had the shape of a hollow object. This is shown on the right side of Fig. \ref{fig:DeformInfl}\,a) for computer simulated data. However, for relative variations in the extensions of test particle $r$ and mask $r_o$ below $\unit[25]{\%}$ a localization was still possible. This was also true if the particles were no longer spherical but rather were deformed to ellipsoids. Moreover, any broadening of the particle representations in the convolved image affected predominantly those directions where a deformation actually was present. This directional broadening could be used to detect deformations of spherical particles and extract their orientations.

\subsection{Deformation analysis of hollow particles}
\begin{figure*}
\centering
\includegraphics[width=0.47\textwidth]{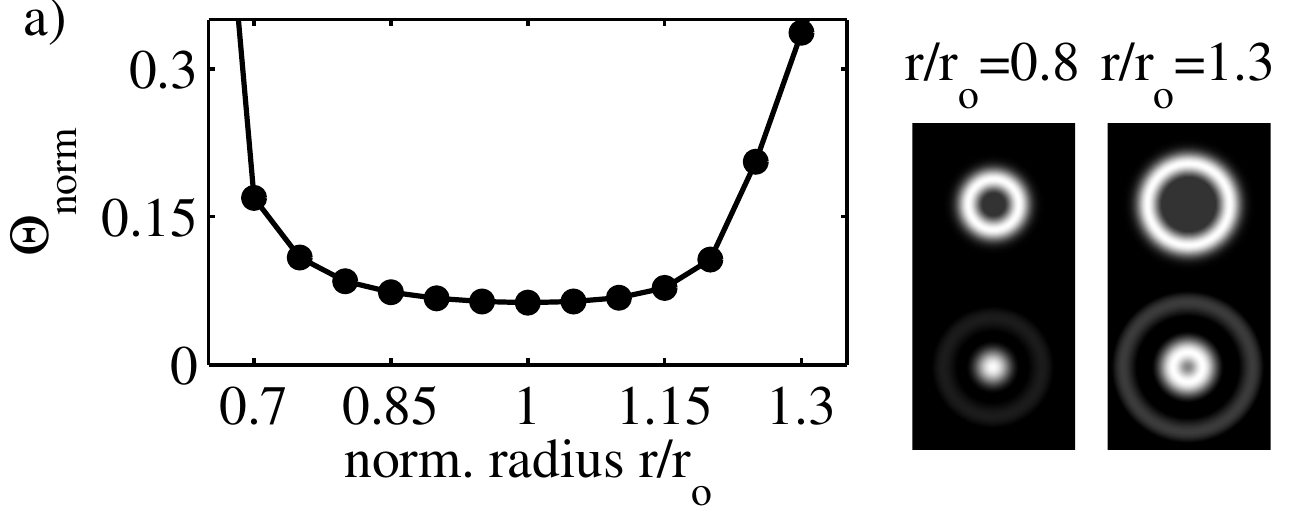}
\hspace{0.03\textwidth}
\includegraphics[width=0.47\textwidth]{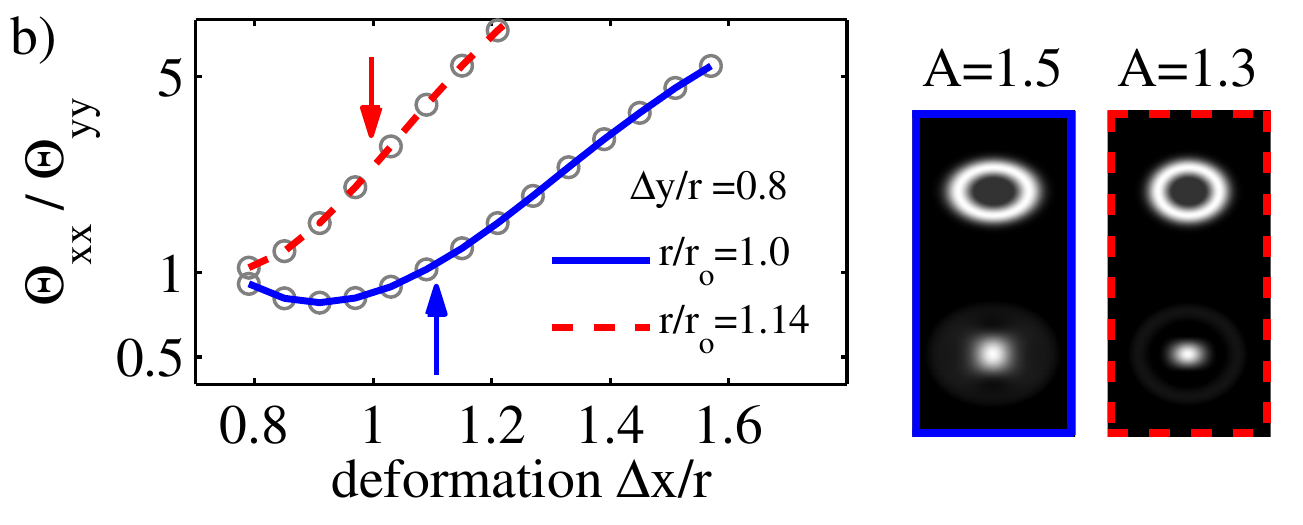}
\caption{a) Deviations in the particle extension of the measured intensity $r$ from the modeled mask $r_o$ lead to a broadening of the convolved image expressed in the increase of $\Theta_\text{norm}$. Two xy-slices of the computer generated 3D particle images (top) and the convolution (bottom) are shown on the right. If the relative deviation are too large $r/r_o \geq 1.3$ the convolution intensity is no longer maximal in the center. b) Deformed particles appear as filled ellipses in the convolved image. The eccentricity of these ellipses is quantified by $\Theta_{xx}/\Theta_{yy}$. This ratio is shown for varying deformations $\Delta x/r$ in x-direction at a constant deformation in y-direction $\Delta y/r$. For matching mask and particle distributions $r/r_o=1$, $\Theta_{xx}/\Theta_{yy}$ does not increase monotonously with increasing aspect ratio $A=\Delta x/\Delta y$. This ambiguity can be resolved via a mismatch $r/r_o\not= 1$. The arrows indicate parameter configuration for which real (top) and convolved (bottom) images are shown on the right.}
\label{fig:DeformInfl}
\end{figure*}
After the localization of the hollow particles the deformation of each particle was extracted individually in a two-step process. In the first step its orientation with respect to the laboratory coordinate frame was analyzed. Afterwards the actual extension of the particle was determined from the original confocal image.

As already mentioned above deformed particles gave rise to an anisotropic convolution signal. This signal resembled the shape of a filled ellipsoid. The major axes of this ellipsoid corresponded to the main directions of the deformation and thus could be used to define the orientation of the deformed hollow particle. For an automated determination of these major axes we introduced the intensity weighted inertia tensor $\Theta_{ab}$.
\begin{eqnarray}
\Theta_{ab}=\frac{1}{\sum_p I_p}\cdot\sum_p I_p\cdot (a_p-\bar{a})\cdot (b_p-\bar{b})
\end{eqnarray} 
with $a, b$ being one of the directions $x,y,z$ and $\bar{a}, \bar{b}$ as the intensity weighted mean positions.  
\begin{eqnarray}
\bar{a}=\frac{1}{\sum_p I_p}\cdot\sum_p I_p\cdot a_p
\end{eqnarray} 
The summations include all pixels $p$ whose convolution intensity $I_p$ exceed a pre-defined threshold value and lie in the region where the particle is localized. The definition of $\Theta_{ab}$ is analogous to the definition of the inertia tensor in classical mechanics with the mass replaced by the convolution intensity. By definition $\Theta_{ab}$ is a symmetric matrix that can be transformed to its diagonal form $\Theta^\prime_{ab}$. The principal axes $\hat{\textbf{u}}_m$ ($m=1,2,3$) obtained from this mathematical procedure can be directly identified with the major axes of the ellipsoid \cite{goldstein1980classical} and thus defines the orientation of the deformed hollow particle. Yet, this procedure relies on the assumption that the particle deformation is unambiguously transferred into the broadening of the convolution signal. As shown in Fig.\,\ref{fig:DeformInfl}\,a) this was not always the case. $\Theta_{ab}$ was calculated for the above discussed variation of the particle radii $r$ at a fixed mask radius $r_o$. As the particles were still spherical in the convolved image [see right in Fig.\,\ref{fig:DeformInfl}\,a)] $\Theta_{ab}$ was already diagonal with equal elements $\Theta_{xx},\,\Theta_{yy}$ and $\Theta_{zz}$. Normalized to the corresponding tensor element of the mask distribution $g_\text{mask}$, $\Theta^\prime_\text{norm}$ was minimal for $r/r_o=1$ and rose irrespective of whether the particle was smaller or larger than the mask distribution. Therefore, a particle that is compressed in one direction and stretched in the other can also lead to a spherical convolution signal for which the orientation cannot be extracted.\\
This effect is demonstrated in Fig.\,\ref{fig:DeformInfl}\,b). An initially spherical particle with radius $r$ was varied regarding its x-deformation $\Delta x/r$ at a fixed compression in y-direction $\Delta y/r=0.8$. Here $\Delta x$ and $\Delta y$ denote the particle extensions in x and y direction, respectively. After the calculation of $\Theta_{ab}$ the ratio of tensor elements $\Theta_{xx}/\Theta_{yy}$ gave a measure of the eccentricity of the convolution signal. If the eccentricity deviated from $1$ the particle orientation could be determined. The mask distribution was chosen to exactly fit the original particle size $r/r_o=1$. As expected  $\Theta_{xx}/\Theta_{yy}$ equaled 1 for equal deformation $\Delta x/r =\Delta y/r=0.8$. But for a x-deformation of about $\Delta x/r=1.1$ corresponding to an aspect ratio of $A=\Delta x/\Delta y=1.5$ the tensor elements evened out again. Hence no orientation could be extracted from the convolved image although it was clearly visible in the initial image [see right in Fig.\,\ref{fig:DeformInfl}\,b)]. \\
In order to resolve this ambiguity the mask distribution properties were chosen to on purpose mismatch the particle properties. This was tested in Fig.\,\ref{fig:DeformInfl}\,b) for a size mismatch of $r/r_o=1.14$. The deformations $\Delta y/r$ and variations in $\Delta x/r$ were kept fixed for better comparison. For this particular set of parameters the eccentricity $\Theta_{xx}/\Theta_{yy}$ rose monotonically with increasing x-deformation allowing for an unambiguous determination of the particle orientation. As a rule of thumb the mask extension should be reduced to the smallest extension of the deformed particles. A variation of other mask distribution properties like the shell thickness altered the slope of the curves in Fig.\,\ref{fig:DeformInfl}\,b) and hence may be used to adjust the sensitivity in aspect ratios. Without further discussion we state that a orientation analysis is possible for moderate deformations. Although the actual numbers depend on the confocal resolution, spatial sampling rate and particle properties, variations in the aspect ratio and deformation of up to $\unit[25]{\%}$ can be handled without any restriction. 
\begin{figure*}
\centering
\includegraphics[width=0.47\textwidth]{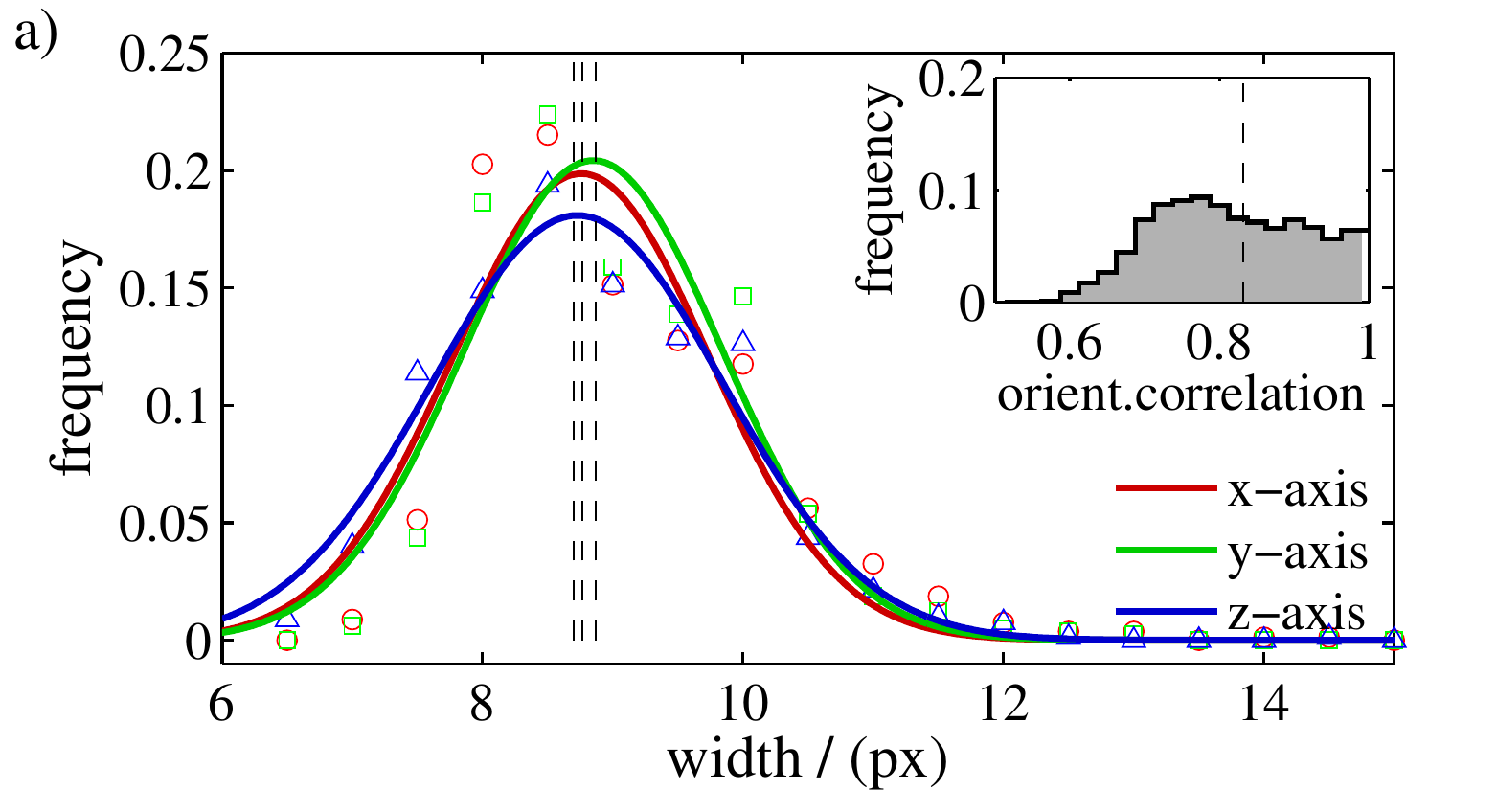}
\hspace{0.03\textwidth}
\includegraphics[width=0.47\textwidth]{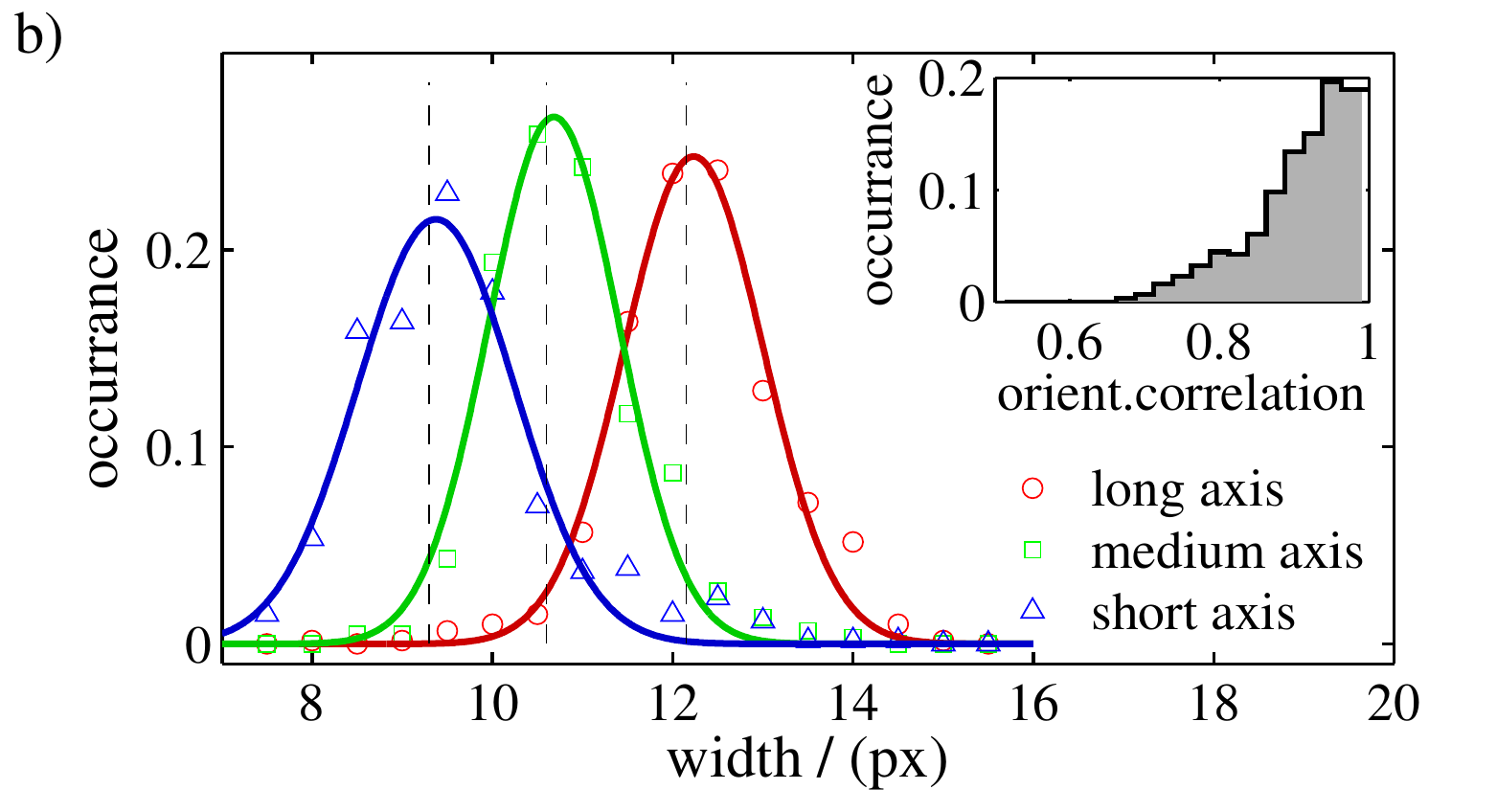}
\caption{a) Normalized distribution of particle extensions for spherical particles. The extracted principal axes lead to identical mean particle extensions (dashed lines). As expected the orientational correlation with the outer coordinate system is flat and is centered around the theoretical value of $0.83$ (dashed line in the inset). b) The analysis of the stretched and rotated data set reproduced the stretching factors of 1, 1.15 and 1.3. The main axes correlate with the rotated coordinate system axes (see inset).}
\label{fig:DeformAna}
\end{figure*}

Once the orientation of the deformed particle is known the actual deformation along the principal axes can be extracted. However, this cannot be done on basis of the diagonal tensor elements $\Theta^\prime_{aa}$ because these quantities are mathematically not fully decoupled. Instead the deformation is measured from the original confocal images. The projection of the intensity distribution along each of the principal axes shows two distinct maxima resulting from the particle shell. The separation distance of these two maxima is a direct measure of the particle extension and also its deformation.

The complete orientation and deformation analysis was tested for a real confocal data set of undeformed hollow spheres. More than $\unit[99]{\%}$ of all particles, about 1000 in total, were located with a sub-pixel accuracy ($\leq \unit[5]{\%}$ of radius). After extraction of the principal axes of $\Theta_{ab}$ the particles extensions were measured in the original unfiltered image giving the same mean value for all three directions [see Fig.\,\ref{fig:DeformAna}\,a)] as it was expected for spherical particles.  
Beside the particle extensions the orientations of the principal axes was tested. Each of the three principal axes $\hat{\textbf{u}}_m$ was assigned to the axes of the laboratory frame $\hat{\textbf{a}}$ with a maximal projection $|\textbf{u}_m \cdot \hat{\textbf{a}}|$. This projection is a measure of the orientational correlation of the principal axes and the laboratory frame. Due to the assignment of axes values below $1/\sqrt{3}\approx 0.58$ do not occur. Besides, the distribution of the projections is flat since there is no preferential direction for the principle axes. Indeed the experimental distribution in the inset in Fig.\,\ref{fig:DeformAna}\,a) was found to be rather flat and its average value agreed reasonably well the theoretical value of $0.83$. 

In order to test the performance of the algorithm for a known deformation and rotation the same data set was stretched virtually along the x- and y-direction with ratios of $\Delta x/r=1.15$ and $\Delta y/r=1.3$, respectively. Additionally, the data set was rotated to exclude a possible influence of the choice of the laboratory system. In the statistical analysis the stretching ratios were reproduced and the principal axes aligned with the axes of the rotated coordinates system giving a maximum in the distributions of the projections at 1 [see Fig.\,\ref{fig:DeformAna}\,b]. 

The only requirement for this analysis to work is a homogeneous fluorescence intensity across the labeled parts of the particle since $\Theta_{ab}$ is sensitive to intensity fluctuations. So, in principle also fully labeled particles can be used. However, here the variations in $\Theta_{ab}$ for deformed particles are less pronounced handicapping the analysis.

At this point we want to discuss some potential difficulties of the analysis that arise from the finite size of the PSF. If the extension of the laser focus largely exceeds the thickness of the labeled shell the measured intensity in the vicinity of two adjacent particles is enhanced. This directly affects the convolved image $C$ as well as the calculation of $\Theta_{ab}$. An efficient solution to this problem is to clip the maximal measured fluorescence intensity by introducing an upper threshold. The situation is getting even worse if the shell thickness is comparable to the size of the PSF. In this case the shells of two adjacent particles merge in the confocal images and the determination of the particle extension after the orientation analysis is hindered. This case should be avoided by an appropriate choice of particle properties and spatial resolutions.\\
The confocal images are also distorted by the anisotropy of the PSF that is broader in z-direction compared to x and y-directions. As a consequence the particle appears blurred in z-directions. We circumvent this problem by blurring the whole 3D fluorescence intensity data in x- and y-directions via an additional convolution with a suitable Gaussian distribution.  

\subsection{Preparation of orientation anisotropic particles via selective bleaching and theoretical polarization contrast}
\label{DEFROT:RotAna:TheoModel}
\begin{figure*}
\centering
\includegraphics[height=4.7cm]{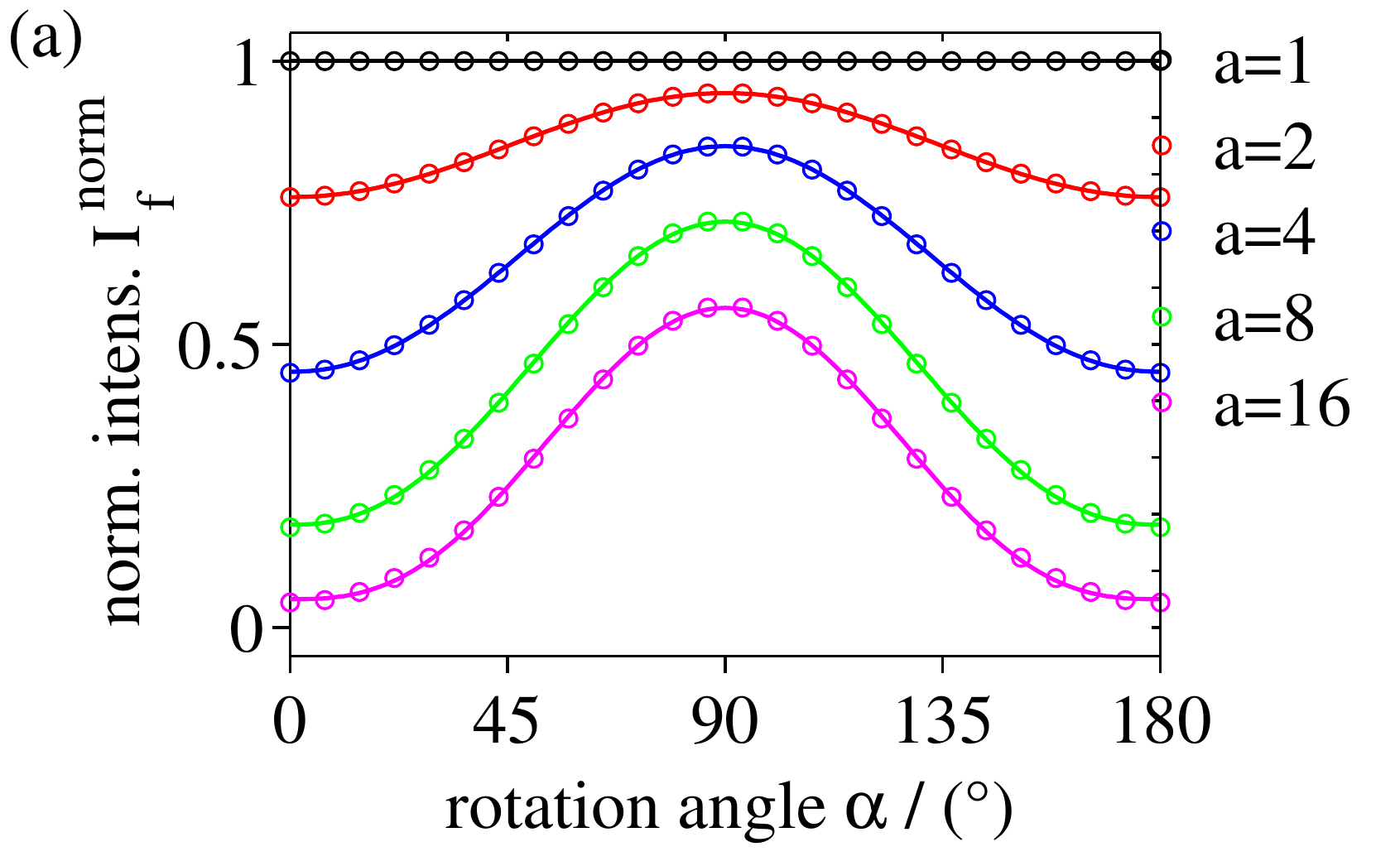}\hspace{0.03\textwidth}
\includegraphics[height=4.7cm]{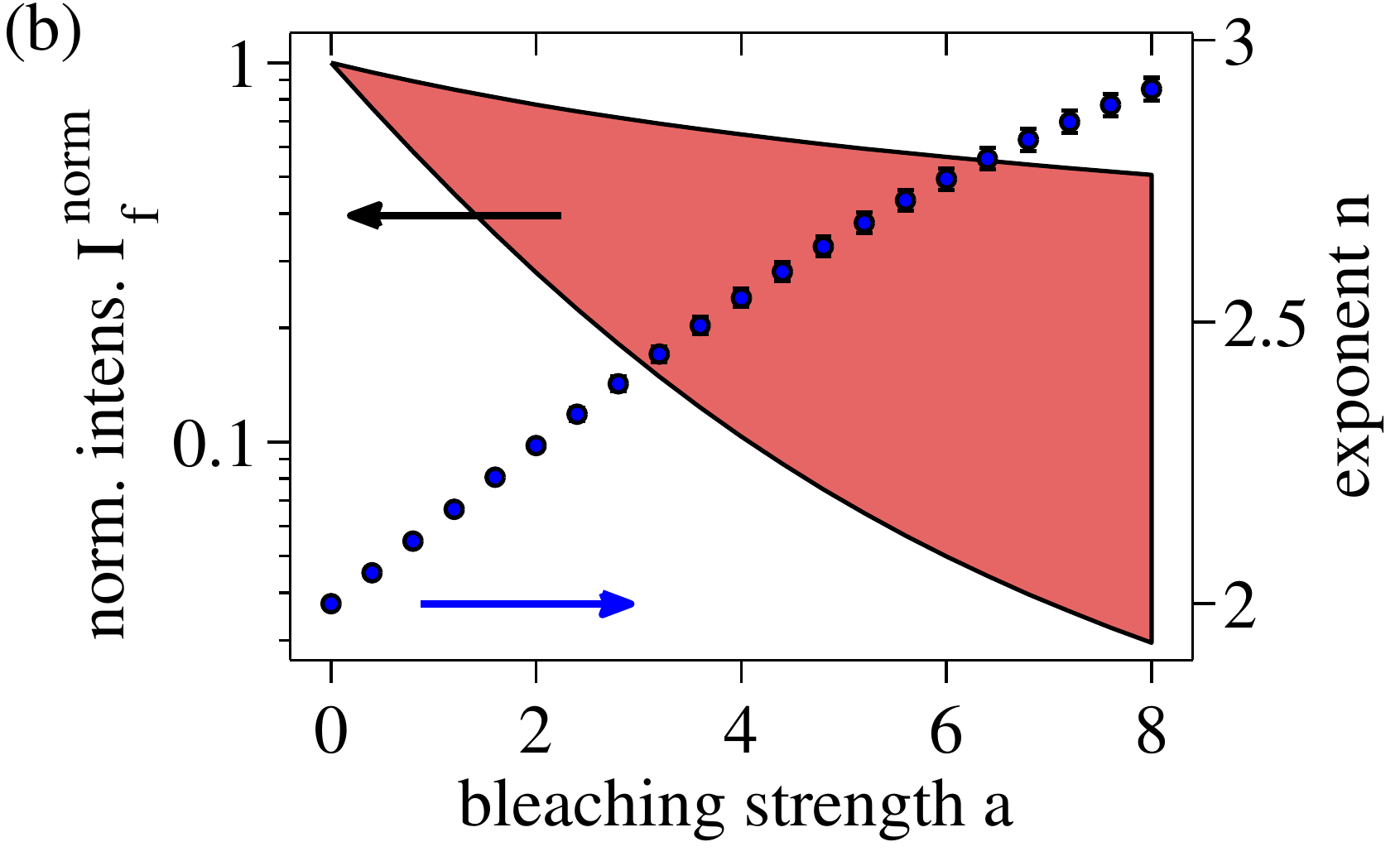}
\caption{(a) Simulated ratio of the fluorescence intensity $I_\text{B}/I_\text{NB}$ of bleached and non-bleached particles as a function of the relative angle between the polarizations of bleach and probe beam for various belaching strengths $a$. The data was fitted by a biased $abs[\sin(\alpha)]^n$ dependency. (b) With increasing bleaching strengths $a$ the polarization contrast $I_\text{B}(\unit[90]{^\circ})/I_\text{B}(\unit[0]{^\circ})$ increased strongly (red shaded area) whereas the exponent $n$ varied moderately.}
\label{fig:BleachSim}
\end{figure*}

In the following we discuss a method to quantify the rotational motion of a fluorescent particle. There are no demands on the shape of the particles but the dye molecules must be immobile, like e.g. rhodamine-B\cite{PCCP.1.4571} that is covalently bound inside the silica particles used in this study. We want to mention that dye molecules that are incorporated into polymeric particles by swelling with a co-solvent cannot be regarded as immobile. Therefore such kind of particles are not suitable. The rotational analysis is based on the fact, that most dye molecules show an approximate dipole characteristic during excitation and emission \cite{JChemPhys.120.4517, Biophys.36.73}.\\
Assuming an isotropic orientation of chromophores before the bleaching process the absorption dipole moment \footnote{We omit any dependencies on the spatial distribution of chromophores within the particles} follows the normalized distribution
\begin{eqnarray}
\boldsymbol{\mu}_\text{A}^{\,\text{init}}(\vartheta,\phi) &=&\dfrac{\mu_\text{A}}{4\,\pi\,\sin(\vartheta)}\,\left[
\begin{array}{c}
\cos(\phi)\,\sin(\vartheta)\\
\sin(\phi)\,\sin(\vartheta)\\
\cos(\vartheta)
\end{array}
\right]
\end{eqnarray}
with $\mu_\text{A}$ denoting the absolute absorption dipole moment and $\vartheta$ and $\phi$ polar and azimuthal angles. In the presence of the electric field
\begin{eqnarray}
\textbf{E}_\text{B}=E_\text{B}\,\left[\begin{array}{c}
1\\0\\0 \end{array}\right]\,\exp[i\,(\omega\,t-k\,z)]
\end{eqnarray}
of a linearly polarized laser beam\footnote{Without restricting generality the polarization of $\textbf E_\text{B}$ was set parallel to the x-direction with a propagation direction of the laser beam along the z-axis} the probability of one molecule to be optically excited is proportional to $|\boldsymbol{\mu}_\text{A}\cdot \textbf{E}_\text{B}|^2$ \cite{JChemPhys.120.4517}. This relation is not only valid for the absorption of photons but also for the bleaching of chromophores if a single photon process is involved \footnote{In most cases the exact bleaching mechanism is not understood \cite{ChinPhysLett.20.1940}. It can be a transition into a metastable non-fluorescent electronic state or a photon-induced chemical conversion of the chromophore. If multi-photon processes are involved the theoretical modeling is more complicated and strongly depends on the incident laser intensity.}. So, the laser beam bleaches predominantly those molecules with orientations parallel to the beam polarization. Moreover, if the bleaching is irreversible the number of chromophores decays exponentially resulting in an anisotropic distribution
\begin{figure}
\centering
\includegraphics[width=0.38\textwidth]{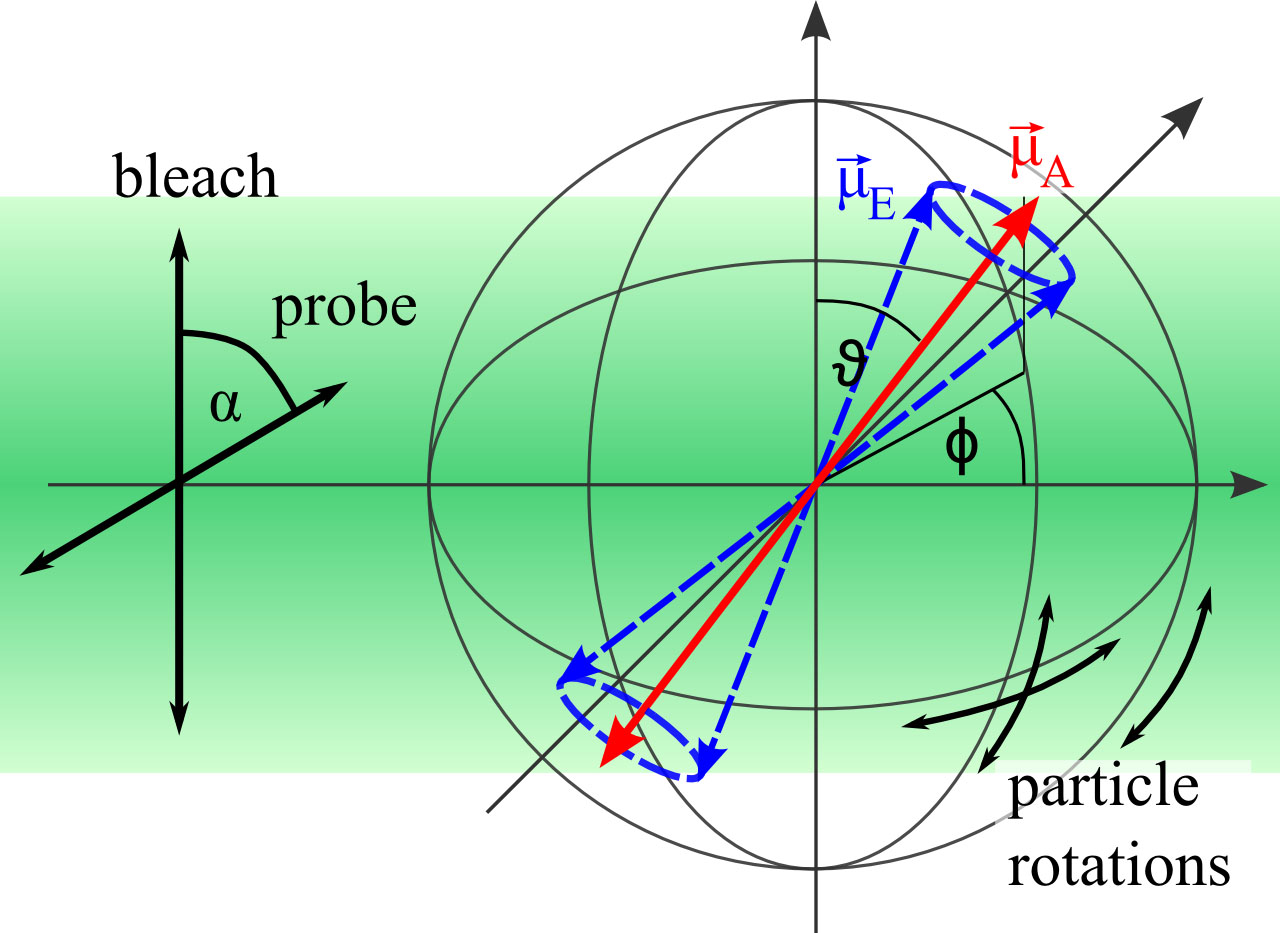}
\caption{Bleaching geometry: The polarizations of bleach and probe beam enclose an angle $\alpha$. The orientation of the absorption dipole $\boldsymbol{\mu}_\text{A}$ of the chromophore is homogeneously distributed on the unit sphere parametrized by polar $\vartheta$ and azimuthal $\phi$ angles. The emission dipole $\boldsymbol{\mu}_\text{E}$encloses a constant angle with $\boldsymbol{\mu}_\text{A}$ that depends on the chromophore. Therefore, all possible orientations of $\boldsymbol{\mu}_\text{E}$ lie on a cone around $\boldsymbol{\mu}_\text{A}$.}
\label{fig:BleachGeometry}
\end{figure}
\begin{eqnarray}
\nonumber
\boldsymbol{\mu}_\text{A}^{\,\text{bleach}}(\vartheta,\phi,a) &=&\boldsymbol{\mu}_\text{A}^{\,\text{init}}(\vartheta,\phi)\cdot\\
&&\qquad\exp\left[-a ( E_\text{B},\mu_\text{A} )\,\cos(\phi)\right]
\end{eqnarray}
The proportionality constant $a$ will be regarded as the bleaching strength and is a linear function in $\mu_\text{A}^2$, $E_B^2$ and the irradiation time $t$. \\
During the subsequent confocal imaging with a probe laser beam 
\begin{eqnarray}
\textbf{E}_\text{P}(\alpha)=E_\text{P}\,\left[\begin{array}{c}
cos(\alpha)\\
\sin(\alpha)\\
0
\end{array}\right]\,\exp[i\,(\omega\,t-k\,z)]
\end{eqnarray} 
enclosing an angle $\alpha$ to the polarization of the bleach beam $\textbf{E}_\text{B}$, the emitted fluorescence intensity is proportional to
\begin{eqnarray}
\nonumber I_f(a) &\sim & \int\limits_{0}^{2\pi}\,\int\limits_{0}^{\pi}|\boldsymbol{\mu}_\text{A}^\text{bleach}(\vartheta,\phi,a)\cdot \textbf{E}_\text{P}(\alpha)|^2\\
&&\qquad\sin(\vartheta)\,d\vartheta\,d\phi
\label{eqn:BleachFluoIntSimple}
\end{eqnarray} 
In the experimental realization the probe beam polarization was adjusted by a polarizer that was placed between the dichroic mirror and the scanning unit in the confocal microscope (\ref{fig:SchematicSetup}). Since the excitation and collected fluorescence light share the same optical path in this part of the setup the fluorescence intensity in \eqref{eqn:BleachFluoIntSimple} is modified by an additional factor $|\boldsymbol{\mu}_\text{E}(\vartheta,\phi)\cdot \textbf E_\text{P}|^2$ whereas $\boldsymbol{\mu}_\text{E}$ denotes the direction of the emission dipole of the chromophore. $\boldsymbol{\mu}_\text{E}$ and $\boldsymbol{\mu}_\text{A}$ are not necessarily parallel. In case of Rhodamine-B both textbftors enclose an angle of about $\unit[23]{^\circ}$ \cite{OptComm.5.307}. This restricts the number of possible orientations of $\boldsymbol{\mu}_\text{E}$ to a cone about $\boldsymbol{\mu}_\text{A}$ (Fig.\,\ref{fig:BleachGeometry}) which is taken into account by an additional integration:
\begin{eqnarray}
\nonumber I_\text{f}(a,\alpha) &\sim & \int\limits_{0}^{2\pi}\,\int\limits_{0}^{\pi}\bigg[\left|\boldsymbol{\mu}_\text{A}^\text{B}(\vartheta,\phi,a)\cdot \textbf{E}_\text{P}(\alpha)\right|^2\cdot \\
\nonumber &&\qquad \int\limits_{0}^{2\,\pi}\left|\left[\mathbf{{R}}(\vartheta,\phi)\,\boldsymbol{\mu}^{\,\text{init}}_\text{A}(\unit[23]{^\circ},\psi )\right] \cdot\textbf{E}_\text{P}(\alpha )\right|^2\,d\psi\bigg]\\
&&\qquad\sin(\vartheta)\,d\vartheta\,d\phi
\label{eqn:BleachFluoIntExt}
\end{eqnarray}   
with the combined rotation matrix
\begin{eqnarray}
\nonumber \left(\mathbf{{R}}(\vartheta,\phi)\right)_{\alpha\beta} &=& \left[
\begin{array}{ccc}
\cos(\vartheta)&0&-\sin(\vartheta)\\
0&1&0\\
\sin(\vartheta)&0&\cos(\vartheta)
\end{array}
\right]\cdot \\
&&\qquad\left[
\begin{array}{ccc}
\cos(\phi)&-\sin(\phi)&0\\
\sin(\phi)&\cos(\phi)&0\\
0&0&1
\end{array}
\right]
\end{eqnarray}
The results of this calculation are illustrated in Fig.\,\ref{fig:BleachSim}\,(a) in dependency on the bleaching strength $a$ and the relative polarization angle $\alpha$. The fluorescence intensity normalized to the initial state before bleaching could be fitted by 
\begin{eqnarray}
I_\text{f}^\text{norm}(a,\alpha)=\dfrac{I_\text{f}(a,\alpha)}{I_\text{f}(0,0)} = I_o - \Delta I\cdot \text{abs}\left[\sin(\alpha)\right]^n
\label{eqn:PolBleachFitFunction}
\end{eqnarray}
The variation of the fit parameters are depicted in Fig.\,\ref{fig:BleachSim}\,(b). Whereas the exponent $n$ varies moderately and is of minor interest, the fluorescence intensity contrast $\Delta I = I_\text{f}^{norm}(a,\unit[0]{^\circ})-I_\text{f}^{norm}(a,\unit[90]{^\circ})$ between bleached  and non-bleached orientations needs to be optimized by choosing an appropriate bleaching strength $a$. If $a$ is too small $\Delta I$ is small and the resolution in the particle rotation is bad. On the other side, if $a$ is too large the fluorescence intensity in the bleached orientation is too small to localize the particle in the confocal image. 

\subsection{Rotational analysis}
\begin{figure*}
\centering
\includegraphics[width=0.9\textwidth]{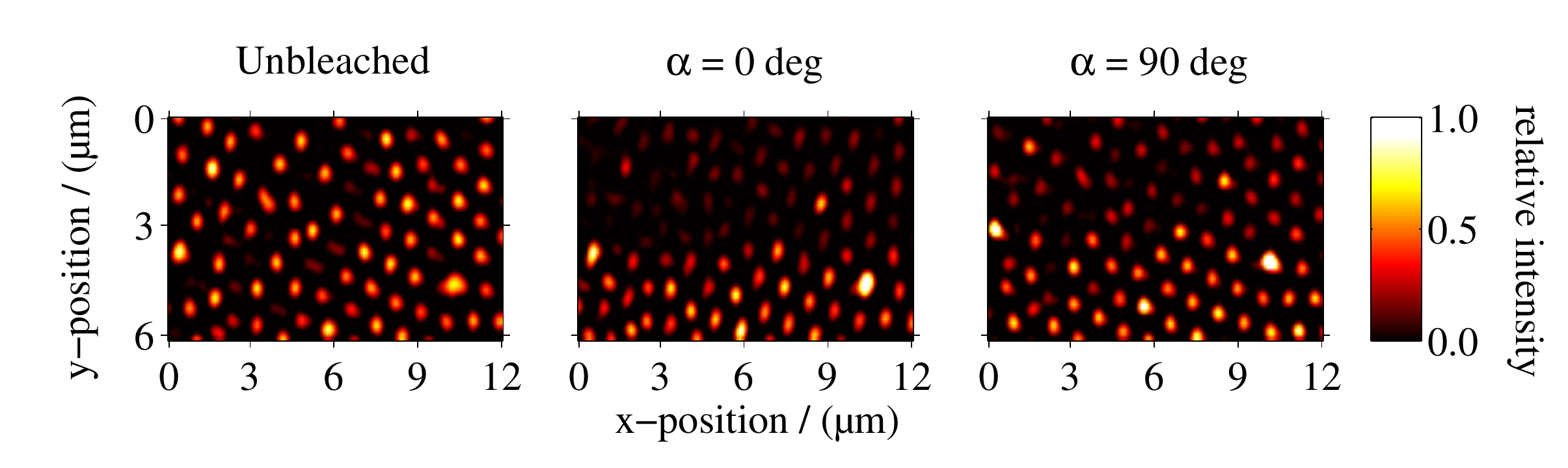}
\caption{Spatially filtered confocal images of a single layer of silica core-shell particles. Before bleaching the measured fluorescence intensity per particle was constant. After bleaching the particles in the upper part of the image the intensity of these particles dropped down by a factor of about $5$ compared to unbleached particles. After changing the relative angle between the polarizations of bleach and probe beam from $\alpha=\unit[0]{^\circ}$ to $\alpha=\unit[90]{^\circ}$ the detected intensity contrast to the unbleached particles decreased.}
\label{fig:BleachSampleData}
\end{figure*}
Instead of rotating particles we demonstrate the usability of this approach by changing the relative polarization of bleaching and probe beam for immobilized particles. Bleaching and confocal imaging was done with the same laser beam enabling for an easy control of the bleaching strength. After selecting a suitable sample area with about 500 particles a fraction of the particles were bleached to a level of about $\unit[30]{\%}$ of the
 initial fluorescence intensity equivalent to a total energy dose of $\unit[0.16]{\mu J/particle}$. After bleaching the laser intensity was reduced to prevent further significant bleaching during imaging of the sample. Fig.\,\ref{fig:BleachSampleData} compares confocal images of the unbleached particles alone and in relation to the bleached particles under relative polarization angles $\alpha$ of $\unit[0]{^\circ}$ and $\unit[90]{^\circ}$. The differences in the fluorescence intensity are clearly visible.
\begin{figure}[t!]
\centering
\includegraphics[width=0.38\textwidth]{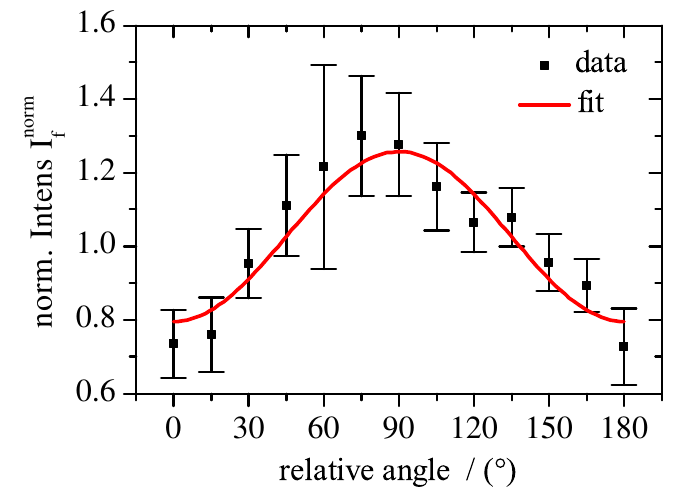}
\caption{Rotation of the laser polarization leads to a variation of the fluorescence intensity that can be fitted by a $\sin^2$ behavior. The intensity is normalized to the average intensity for all angles.}
\label{fig:PolBleach}
\end{figure}
Beside to these two extreme values 3D confocal images were taken for various relative polarization angles. From each image the position and total fluorescence intensity of each bleached particle was extracted and averaged (Fig.\,\ref{fig:PolBleach}). The fluorescence contrast $\Delta I$ amounted to about $\unit[40]{\%}$ and the functional dependency on $\alpha$ could be fitted to equation \eqref{eqn:PolBleachFitFunction} when fixing the exponent to $n=2$. \\
Although the theoretical modeling disregards the strong focusing of bleach and probe laser beam by the objective the dependency on $\alpha$ is reasonably well described. The large statistical scatter in the relative intensities might be an indication of this discrepancy to the model as the divergence of the beam tends to blur the bleaching contrast. 

\section{Conclusions}
\label{sec:Conculsions}
We showed that fluorescent confocal microscopy cannot only be used to obtain 3D coordinates of a particle but at the same time can deliver information about its deformation and rotation. This can be done on a single particle level in a fully automated procedure. Although there are no specific demands on the particles in particular hollow spheres are a good choice for the investigation of deformations in colloidal and granular systems. If the shell is thin compared to the used excitation wavelength ($d\leq\lambda/10$) there is no need for exact matching of the refractive indexes of particle and solvent. Hence, the choice of compatible materials is considerably increased. On the other side rotational and frictional sliding motions of particles have a strong impact on reorganization effects in arrested colloids. The described method can be used to extract this information, even for dense systems of spherical particles.
\begin{acknowledgments}
We thank Pavlik Lettinga and Doris Vollmer for stimulating discussions. M.F., M. d'A. and G.K.A. acknowledges SPP 1273 \textquotedblleft Kolloidverfahrenstechnik\textquotedblright (Au321/1-1,2,3) as well as SPP 1486 \textquotedblleft Partikel im Kontakt\textquotedblright (Au321/2-1) for financial support. M.R. is a recipient of a fellowship through funding of the Excellence Initiative (DFG/GSC 266). 
\end{acknowledgments}

\bibliography{literature}

\end{document}